\documentclass[12pt]{article}

\usepackage{listings}
\usepackage{xcolor}
\usepackage{booktabs}
\usepackage[utf8]{inputenc}
\usepackage[english]{babel}

\usepackage{graphicx,xurl}
\usepackage[version=4]{mhchem}

\usepackage{siunitx}
\usepackage[hidelinks]{hyperref}
\usepackage{sbc-template}

\definecolor{gray}{gray}{0.95}

\lstdefinelanguage{XML}
{
  morestring=[b]",
  morestring=[s]{>}{<},
  morecomment=[s]{<!--}{-->},
  stringstyle=\color{blue},
  identifierstyle=\color{black},
  keywordstyle=\color{black},
  morekeywords={id,value,speed,pstate,prop,host},
}

\title{Estimating \ce{CO2} emissions of distributed applications and platforms with SimGrid/Batsim}

\author{Gabriella Saraiva\inst{1}\and Miguel Vasconcelos\inst{2} Sarita Mazzini Bruschi\inst{3}\and\\ Danilo Carastan-Santos\inst{4}\and Daniel Cordeiro\inst{1}}

\address
{Escola de Artes, Ciências e Humanidades  --- Universidade de São Paulo 
(EACH--USP) --- \\
  São Paulo, SP --- Brasil
  \nextinstitute University of Toulouse, CNRS, Toulouse INP, UT3 --- Toulouse, France 
  \nextinstitute
  Instituto de Ciências Matemáticas e de Computação --- Universidade de São Paulo \\
  (ICMC--USP) --- São Carlos, SP --- Brasil
  \nextinstitute
  University Grenoble Alpes, CNRS, Inria, Grenoble INP, LIG\\
  Grenoble --- France
  \email{gabriella\_saraiva@usp.br, miguel-felipe.silva-vasconcelos@irit.fr}\vspace{-1em}
  \email{sarita@icmc.usp.br, danilo.carastan-dos-santos@univ-grenoble-alpes.fr}\vspace{-1em}
  \email{daniel.cordeiro@usp.br}
}

\begin{document} 

\maketitle

\begin{abstract}
  This work presents a carbon footprint plugin designed to extend the capabilities of the Batsim simulator by allowing the calculation of \ce{CO2} emissions during simulation runs. The goal is to comprehensively assess the environmental impact associated with task and resource management strategies in data centers. The plugin is developed within SimGrid---the underlying simulation framework of Batsim---and computes carbon emissions based on the simulated platform's energy consumption and carbon intensity factor of the simulated machines. Once implemented, it is integrated into Batsim, ensuring compatibility with existing simulation workflows and enabling researchers to assess the carbon efficiency of their scheduling strategies.
\end{abstract}
     
\section{Introduction}

The rapid growth of digitalization across various sectors has increasingly demanded computational infrastructure, turning data centers into major electricity consumers. According to the International Energy Agency, these facilities account for approximately 2\% of global energy consumption \cite{IEA2024}. This is mainly due to the intensification of internet data traffic and the expansion of workloads and storage capacity. In the United States, for example, the energy consumption of data centers increased from 91 billion \unit{\kWh} in 2013 to almost double by 2020, highlighting the impact of this evolution \cite{Abbas2021}.

This increase has raised significant environmental concerns. In response to the energy and climate crisis, countries like China have established sustainability goals, including the commitment to peak carbon emissions by 2030 and achieve carbon neutrality by 2060 \cite{ZHANG2025207}. Meanwhile, the industry has been striving to improve the energy efficiency of data centers. Despite increasing demand, global energy consumption by data centers increased only 6\% between 2010 and 2018, as a result of efforts such as technological modernization, improvements in cooling systems, and advances in workload management algorithms \cite{Masanet2020}.

However, the environmental impact of distributed computing infrastructures goes beyond energy consumption. The carbon footprint of data centers is determined not only by the amount of energy used but also by the carbon intensity of the electricity consumed, which varies significantly by region and over time. For example, Google's 2025 Environmental Report shows that while some of its data centers operate with almost 100\% renewable energy, others, such as those in Japan, rely on grids with only 4\% renewable share \cite{google2025}. Microsoft's sustainability reports also highlight that, as the share of renewables increases, indirect emissions from manufacturing IT equipment become more relevant \cite{microsoft2025}.

Various strategies can be applied to reduce energy consumption in these environments. At the infrastructure level, adopting more efficient components, such as optimized CPUs or specialized accelerators (e.g., GPGPUs), already represents progress. Intelligent task and resource management play a key role at the software level. Among the most effective measures are to reduce task execution times, use available resources efficiently, and even turn off idle machines \cite{Poquet2017}. These decisions are guided mainly by scheduling algorithms that analyze usage patterns to strategically determine where, when, and how each task should be executed, to balance energy efficiency and performance.

\section{Background and Related Work}

Simulators have been widely used in large-scale computing environments to study and test scheduling approaches before practical deployment. They provide a controlled means of modeling the behavior of data centers, allowing the evaluation of different strategies in terms of load distribution, response times, energy consumption, and resource usage. This enables operators to validate complex scenarios and develop solutions that improve platform performance while reducing environmental impact.

Several simulation frameworks have been developed to model and analyze distributed systems' performance and energy consumption. Among the most widely adopted are SimGrid~\cite{simgrid2014}, CloudSim~\cite{cloudsim2011}, and GreenCloud~\cite{greencloud2011}. 

SimGrid is well established within the scientific community and provides comprehensive support for simulating various distributed environments. Incorporating energy consumption and renewable energy generation models, such as solar panels. However, SimGrid does not natively support the calculation of \ce{CO2} emissions based on the carbon intensity of the electricity mix utilized by the simulated infrastructure.

This work introduces a carbon footprint plugin developed in SimGrid and integrated in Batsim~\cite{batsim_jsspp16}, with the objective of enabling a more comprehensive analysis of the environmental impact of different management strategies in distributed applications and platforms. 
Our plugin allows estimating \ce{CO2} emissions associated with energy consumption during simulated executions in Batsim, considering variables such as the carbon intensity of the energy source used. Our plugin may be useful for researchers and practitioners to assess not only the energy efficiency of their solutions but also their environmental impacts, contributing to the development of more sustainable technologies in the context of high-performance computing.
The source code for the plugin, as well as the documentation and tutorials, is openly available\footnote{Plugin Home Page: \url{https://github.com/saraiva03/carbon-footprint-simgrid-batsim}}

\section{Carbon Footprint plugin}

The carbon footprint plugin\footnote{Carbon footprint plugin repository: \url{https://github.com/saraiva03/simgrid/tree/carbon-footprint-calc}} for SimGrid was designed to enable simulation and monitoring of \ce{CO2} emissions associated with the energy consumption of computational hosts in distributed systems. The plugin provides a flexible and extensible framework that integrates seamlessly with SimGrid's energy plugin, allowing users to quantify the environmental impact of their simulated workloads based on customizable carbon emission intensities.

\subsection{Mathematical Model}

An incremental model based on the energy consumption of each host was implemented to estimate the carbon footprint associated with computational workloads. This work applies the model exclusively to the machines (hosts) in the simulated environment. The energy consumption of network equipment and cooling systems is not considered at this stage and may be addressed in future work.

The model accounts for both the static component (energy consumed when the machine is idle) and the dynamic component (additional energy consumed when the machine executes tasks). Carbon emissions are periodically updated during the simulation, using the host's instantaneous power consumption and the environment's carbon intensity.

Let $t_0$ be the time of the last update and $t_1$ the current simulation time. At each update, the instantaneous power consumption of the host, $P$, is measured at time $t_0$ and is assumed to remain constant during the interval $[t_0, t_1]$. Therefore, the energy consumed is given by:

\begin{equation}
    E_{\text{step}} = P \cdot (t_1 - t_0)
\end{equation}

\begin{equation}
    E_{\text{step}}^{\unit{\kWh}} = \frac{E_{\text{step}}}{3.6 \times 10^6}
\end{equation}

\begin{equation}
    C_{\text{step}} = E_{\text{step}}^{\unit{\kWh}} \cdot CI_{\text{step}}
\end{equation}

\begin{equation}
    C_{\text{total}}(t_1) = C_{\text{total}}(t_0) + C_{\text{step}}
\end{equation}

Where:
\begin{itemize}
    \item $E_{\text{step}}$: energy consumed in the interval (in joules);
    \item $E_{\text{step}}^{\unit{\kWh}}$: energy in kilowatt-hours;
    \item $P$: instantaneous power consumption of the host at the time $t_0$ (in watts)
    \item $CI_{\text{step}}$: carbon intensity of the environment in the interval (in \unit{\g\per\kWh});
    \item $C_{\text{step}}$: carbon emitted in the interval (in grams);
    \item $C_{\text{total}}$: total carbon footprint up to time $t_1$.
\end{itemize}
\vspace{+0.2em}

This model is implemented in the \texttt{HostCarbonFootprint::update()} function, ensuring that the estimated carbon emissions reflect the host's dynamic behavior during the simulation. The carbon intensity $CI$ can be configured according to the energy profile of the simulated environment.

\subsection{Overview of the Plugin Functionality}

The carbon footprint plugin was developed to extend SimGrid's host model, enabling the quantification of carbon emissions associated with energy consumption during simulation. An extension is attached to each \texttt{Host} object, responsible for monitoring and updating the accumulated emissions according to the host's activity to achieve this.

The plugin operates by intercepting host key simulation events, such as initialization (creation), shutdown (destruction), state transitions (power on/off), and changes in execution profile (task execution or updates to carbon intensity). Whenever one of these events occurs, the plugin updates the host's carbon footprint, considering the energy consumed since the last update and the current carbon intensity at that moment.

The carbon intensity, defined as the amount of \ce{CO2} emitted per unit of energy consumed (\unit{\g\per\kWh}), can be individually configured for each host via properties in the platform XML file, as illustrated in Listing~\ref{lst:xml-host2}. Furthermore, the plugin allows for the dynamic adjustment of this parameter during simulation, reflecting possible changes in the energy profile of the simulated environment.
\vspace{+0.5em}

\begin{figure}[ht]
    \centering
    \includegraphics[width=0.7\linewidth]{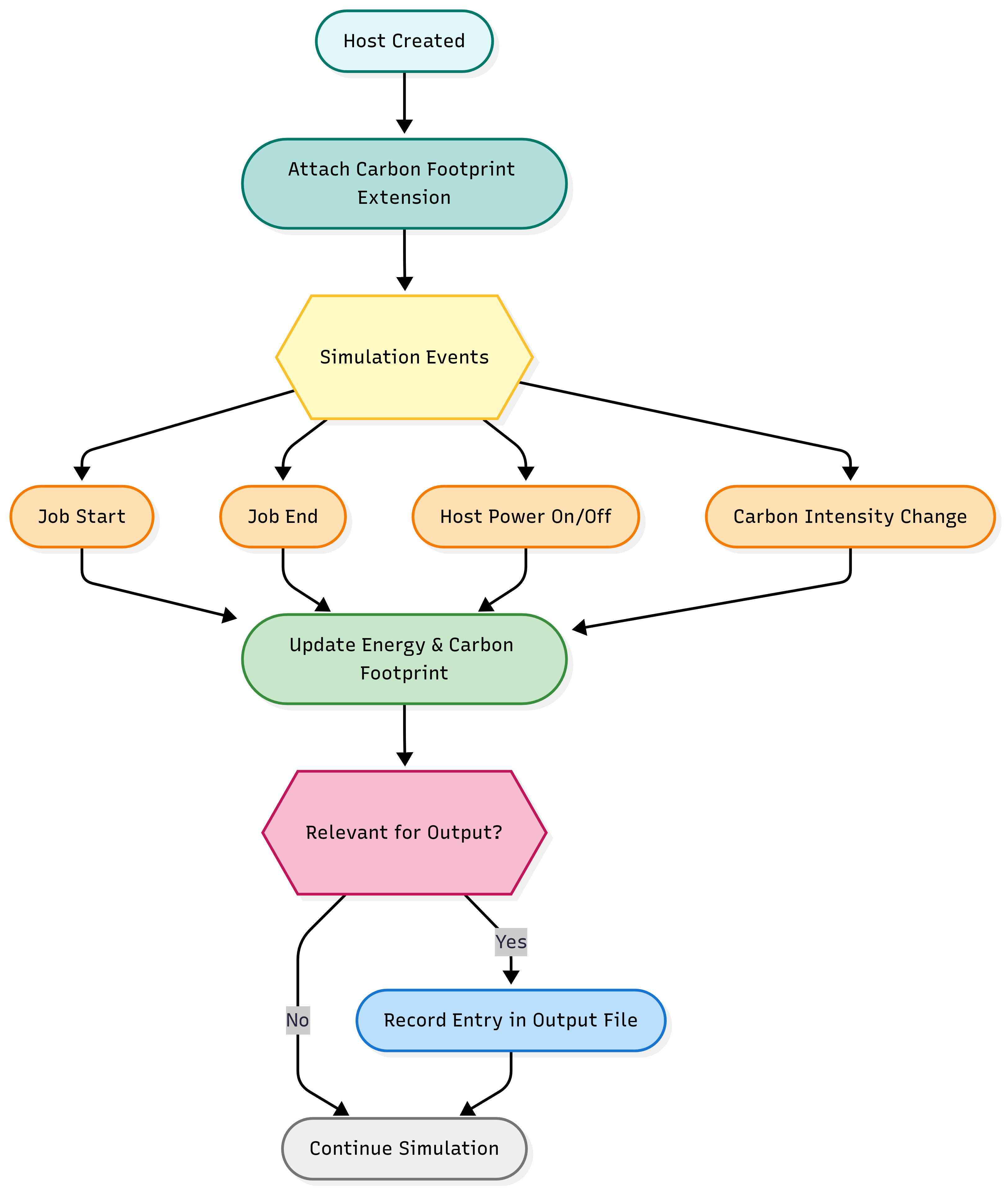}
    \caption{Flowchart illustrating the main steps of the carbon footprint plugin and its integration with Batsim.}
    \label{fig:carbon-plugin-flow}
\end{figure}

SimGrid was modified to include callbacks on the main host events to enable this functionality, ensuring that the carbon footprint extension is notified and can perform the necessary updates. As a result, the original SimGrid model is enhanced to consider not only energy consumption but also the environmental impact of simulated executions, providing detailed metrics on carbon emissions over time. The overall workflow of the plugin is illustrated in Figure~\ref{fig:carbon-plugin-flow}.

\subsubsection{Validation}
The plugin was validated using different test case scenarios to explore the different states in SimGrid regarding the energy consumption of machines: i) when the machine is off; ii) when the machine is idle (the machine is powered on but not running any computations); and iii) when the machine performs computations, using one or more of its CPU cores.

Two main scenarios were considered for the validation regarding the carbon emissions. First, we use a static value for the carbon emissions of the electricity grid, which can represent countries for which we only have the annual value of the carbon intensity. The second scenario considers values of carbon intensity that change over time, for example, to represent countries with intermittent renewable sources in their mix. 

\begin{figure}[ht]
\centering    \includegraphics[width=.5\textwidth]{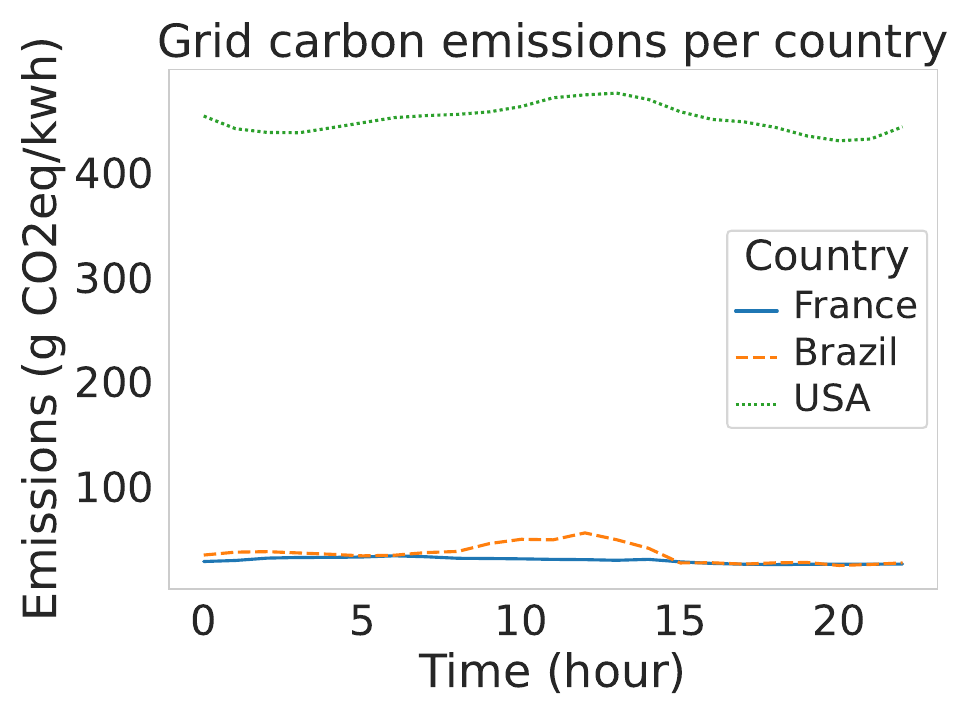}
    \caption{Grid emissions values for the different electricity mixes considered.}
    \label{fig:emissions_trace}
\end{figure}

Figure~\ref{fig:emissions_trace} illustrates the variability over a single day for the different carbon intensities in the different countries we considered: i) the USA, which shows a high carbon intensity because it uses high-carbon sources such as coal; ii) France, which shows a low carbon intensity given the high presence of nuclear power; and iii) Brazil, which also shows a low carbon intensity, given the presence of renewable sources such as hydroelectric power.

The validation step is important to ensure that the plugin remains working in the future when new code is incorporated into the SimGrid framework, such as including new functionalities or fixes to other problems in the code.
Further details about the test scenarios and the inputs used can be found in the plugin's GitHub repository.

\subsection{Integration with Batsim}

The carbon footprint plugin was developed to monitor carbon emissions during simulations in Batsim. To use it, add the ``\texttt{-C}'' or ``\texttt{--carbon-footprint}'' option when running Batsim, along with the platform and workload files. This activates the plugin, which then begins tracking each host's energy consumption and carbon emissions in the simulation.

During execution, Batsim monitors key events such as the start and end of jobs, as well as changes in the power state of the hosts. Whenever one of these events occurs, the updated carbon emission value is retrieved for each host and recorded in an output file. This file contains information such as the event timestamp, the energy consumed, the total carbon emissions, the type of event, and the average emissions since the last recorded entry.
The source code for the version of Batsim with the integrated carbon footprint plugin is available at \url{https://github.com/saraiva03/batsim/tree/carbon-footprint-calc}.

\paragraph{Output Formats}
Once integrated into Batsim, the plugin monitors and logs the environmental impact of scheduling decisions. This is achieved through a tracer mechanism that is triggered on job start, job completion, and host power state changes.

Whenever one of these events is detected, Batsim retrieves the current carbon footprint of the affected host and appends a new entry to the output file. Each entry includes the following fields:
\begin{description}
\item[\texttt{time}] The simulation timestamp at which the event occurred, measured from the beginning of the simulation.
\item[\texttt{energy}] The total energy consumed by the platform (in joules) up to the recorded time.
\item[\texttt{carbon\_emission}] The cumulative carbon emissions (in grams per kilowatt-hour) of all hosts up to that moment.
\item[\texttt{event\_type}] The nature of the event:
\begin{itemize}
\item \texttt{s}: job start
\item \texttt{e}: job end
\item \texttt{p}: power state change
\end{itemize}
\item[\texttt{ecarbon}] The average carbon emission of the platform during the interval between the previous event and the current one.
\end{description}
This structured output allows for detailed temporal analysis of both energy consumption and carbon emissions throughout the simulation.

\section{Evaluation and Comparison}

The experiments aimed to compare the energy consumption and carbon emissions data between two distinct approaches: the actual execution of the inference benchmarks using machine learning models and the simulation of these same applications using Batsim. The implementation developed for the experiments can be accessed at \url{https://github.com/saraiva03/carbon-plugin-tests}.

The real executions involved 100 inferences for each model: ResNet18, BERT large (fine-tuned on the SQuAD dataset), and Deep Learning Recommendation Model (DLRM). These tests were performed on a Dell G15 laptop with an 11th Gen Intel Core i5-11400H processor, featuring six physical cores and a base frequency of 2.70 GHz. The CodeCarbon\footnote{CodeCarbon repository: \url{https://github.com/mlco2/codecarbon}} tool was used to monitor energy consumption, hardware usage, and environmental impact.
The simulated versions of these applications were executed in Batsim, using a configuration model designed to represent the same hardware environment as the real executions.
For both approaches---real and simulated---10 runs were performed. %
The following sections detail the experimental setup and analyze the results.

\subsection{Experimental Setup}

The experiments included several important considerations and assumptions. First, the \texttt{carbon\_intensity} value was calculated as the ratio between carbon emissions (\texttt{emissions}) and energy consumed (\texttt{energy\_consumed}), both extracted from the output file generated by CodeCarbon. This calculation yielded a value of \qty{98.348}{\g\per\kWh}, consistent with the default value defined in the official CodeCarbon documentation. This value reflects the global average carbon intensity of electricity, as presented in the CodeCarbon repository, while considering country-specific variations. In this study, Brazil was used as the reference country.

To ensure accuracy in estimating energy consumption, the \texttt{cpu\_energy} metric was used instead of \texttt{energy\_consumed}, since the real benchmarks were executed exclusively on the CPU of the designated machine. CodeCarbon enables a breakdown of energy usage by hardware components (such as CPU and GPU); therefore, only CPU-specific consumption was considered. Consequently, the Batsim simulation environment was configured to reflect the exact CPU specifications, as it was the only component involved in performing the inference tasks.

Throughout real executions, CodeCarbon recorded CPU power consumption ranging from \qtyrange{30}{40}{\watt}, with occasional peaks reaching \qty{50}{\watt}. This consumption range corresponds to the values reported in the official processor documentation\footnote{Intel's official specification for the processor used (Intel® Core™ i5-11400H): \url{https://www.intel.com.br/content/www/br/pt/products/sku/213805/intel-core-i511400h-processor-12m-cache-up-to-4-50-ghz/specifications.html}} and was used to define the different CPU power states in Batsim.

The number of floating point operations (FLOP) per inference was estimated using the THOP (PyTorch-OpCounter)\footnote{THOP (PyTorch-OpCounter) repository: \url{https://github.com/Lyken17/pytorch-OpCounter}} library, which is built on PyTorch. This library calculates the theoretical computational cost of a model's architecture based on input dimensions and model structure, independent of the actual input data. Therefore, the number of FLOP per inference remained constant across all runs, resulting in minimal variation in energy consumption and carbon emissions recorded during the 10 Batsim simulations.

For comparative analysis between real and simulated executions, standard statistical metrics were used, including the R\textsuperscript{2} score, Mean Absolute Percentage Error (MAPE), and Root Mean Square Error (RMSE).

\textbf{Machine Description and Simulator Parameters}
\vspace{+0.5em}

In the Batsim simulator, the Intel Core i5-11400H processor was represented as a host with the following parameters:
\vspace{+1.5em}

\begin{lstlisting}[caption=XML Configuration of Host used in the experiments, label=lst:xml-host2]
<!-- Intel_i5_11400H -->
<host id="Intel_i5_11400H" speed="XGf" pstate="0" core="6">
  <prop id="wattage_per_state" value="10:25:40" />
  <prop id="wattage_off" value="1.0" />
  <prop id="carbon_intensity" value="98.348" />
</host>
\end{lstlisting}

\begin{description}
  \item \textbf{Host ID (\texttt{id}):} Intel\_i5\_11400H
  \item \textbf{Speed (\texttt{speed}):} Represents the processing capacity per core (in GigaFLOP per second).
  \item \textbf{Number of cores (\texttt{core}):} 6
  \item \textbf{Power states (\texttt{wattage\_per\_state}):} Indicates energy consumption (in watts) at different usage levels:
  \begin{itemize}
    \item \textbf{Idle:} When the processor is idle or under minimal activity, it consumes little energy.
    \item \textbf{Epsilon:} Represents an intermediate load, where some cores are in use or there is moderate activity.
    \item \textbf{AllCores:} Maximum usage state, with all cores active simultaneously, resulting in the highest energy consumption.
  \end{itemize}
  
  \item \textbf{Power consumption in off state (\texttt{wattage\_off}):} \qty{1.0}{\watt}
  
  \item \textbf{Carbon intensity (\texttt{carbon\_intensity}):} \qty{98.348}{\g\per\kWh}, the same value used in the real experiment, as specified in the CodeCarbon documentation.
\end{description}

The speed value was estimated by dividing the total FLOP required for each benchmark by the average execution time, resulting in a performance rate in GFLOP per second per core. Meanwhile, the values assigned to \texttt{wattage\_per\_state} were defined based on measurements of CPU energy consumption, obtained using the Intel Power Gadget software.

\subsection{Results and Analysis}

This section presents the results of the experiments, comparing the energy consumption and carbon emissions data obtained from real executions with the values generated by the Batsim simulation.  The results, extracted from the \texttt{collected\_results.csv} dataset, are analyzed using quantitative metrics, including the R\textsuperscript{2} score, RMSE, and MAPE. The discussion is further supported by a visual analysis of the boxplots, illustrating the distribution and variability of the data for each benchmark.

\textbf{ResNet18: }
The ResNet18 benchmark involved a total of 182 GFLOP across 100 inferences. With a total execution time of approximately 2.5 seconds, the CPU's effective processing rate was 72.8 GFLOP/s, or 12 GFLOP/s per core ($72.8 \div 6$), which was used as the speed value for the Batsim simulation. The simulation results for this model showed an R\textsuperscript{2} score of -0.580, an RMSE of 0.001, and a MAPE of 23.11\%.

\textbf{BERT Large (fine-tuned for SQuAD): }
The BERT Large benchmark involved a total of 1,250 GFLOP across 100 inferences, completed in about 46 seconds. The CPU demonstrated an effective processing rate of 27 GFLOP/s, or approximately 4.5 GFLOP/s per core ($27 \div 6$), the speed value used in the simulator. The simulation of this model produced an R\textsuperscript{2} score of -0.146, an RMSE of 0.004, and a notably lower MAPE of 8.32\%.

\textbf{Deep Learning Recommendation Model (DLRM): }
The DLRM benchmark, the lightest model tested, required 13 GFLOP for 100 inferences, completed in approximately 0.45 seconds. The effective processing rate was 28.8 GFLOP/s, or 4.8 GFLOP/s per core ($28.8 \div 6$). The simulation metrics for DLRM were an R\textsuperscript{2} score of -0.019, an RMSE of 0.000, and a MAPE of 16.13\%, an intermediate error value.

Analyzing the results obtained in the experiments allows us to assess the accuracy of the simulation performed in Batsim compared to the real executions monitored by CodeCarbon. Overall, the simulation was able to estimate the average energy consumption and carbon emissions reasonably close to the real values, especially for heavier models such as BERT Large. However, there are important differences in the variation of the results, reflected in the statistical metrics calculated.

For the ResNet18 model, the mean absolute percentage error (MAPE) was approximately 23.11\%, indicating a significant deviation between the real and simulated values.
However, it is important to note that the power consumption of processors of the same model can vary up to 20\% due to variability in manufacturing~\cite{chasapis2019power}. A MAPE of 23.11\% is therefore close to an acceptable range.
Furthermore, the coefficient of determination (R\textsuperscript{2}) showed a negative value (-0.580), revealing that the simulation could not reproduce the fluctuations observed in real executions. This behavior can be explained by the fact that the model is relatively lightweight and runs quickly, making it more susceptible to consumption variations caused by background processes and dynamic frequency adjustments in the processor. In contrast, the Batsim simulation maintained stable values that did not reflect these occasional consumption peaks.

In the case of BERT Large, the results were more consistent. The MAPE was only 8.32\%, demonstrating a reasonable proximity between the simulated and experimentally measured values. Although it remained negative (-0.146), indicating a low correlation with the variations in different runs, the average consumption and emissions estimated by Batsim were quite accurate. This suggests that for heavier applications with longer execution times, the simulation model represents the energy behavior of the system in a more reliable way.

Finally, for the DLRM model, the MAPE recorded was 16.13\%, an intermediate error compared to the other two benchmarks. Like ResNet18, R\textsuperscript{2} was practically zero (-0.019), again evidencing the difficulty in capturing variations in fast executions. This result is consistent considering that execution times less than one second are more sensitive to measurement inaccuracies and fluctuations in the real system, which reduces the correlation with the simulated values.

\begin{figure}[ht]
\centering    \includegraphics[width=.8\textwidth]{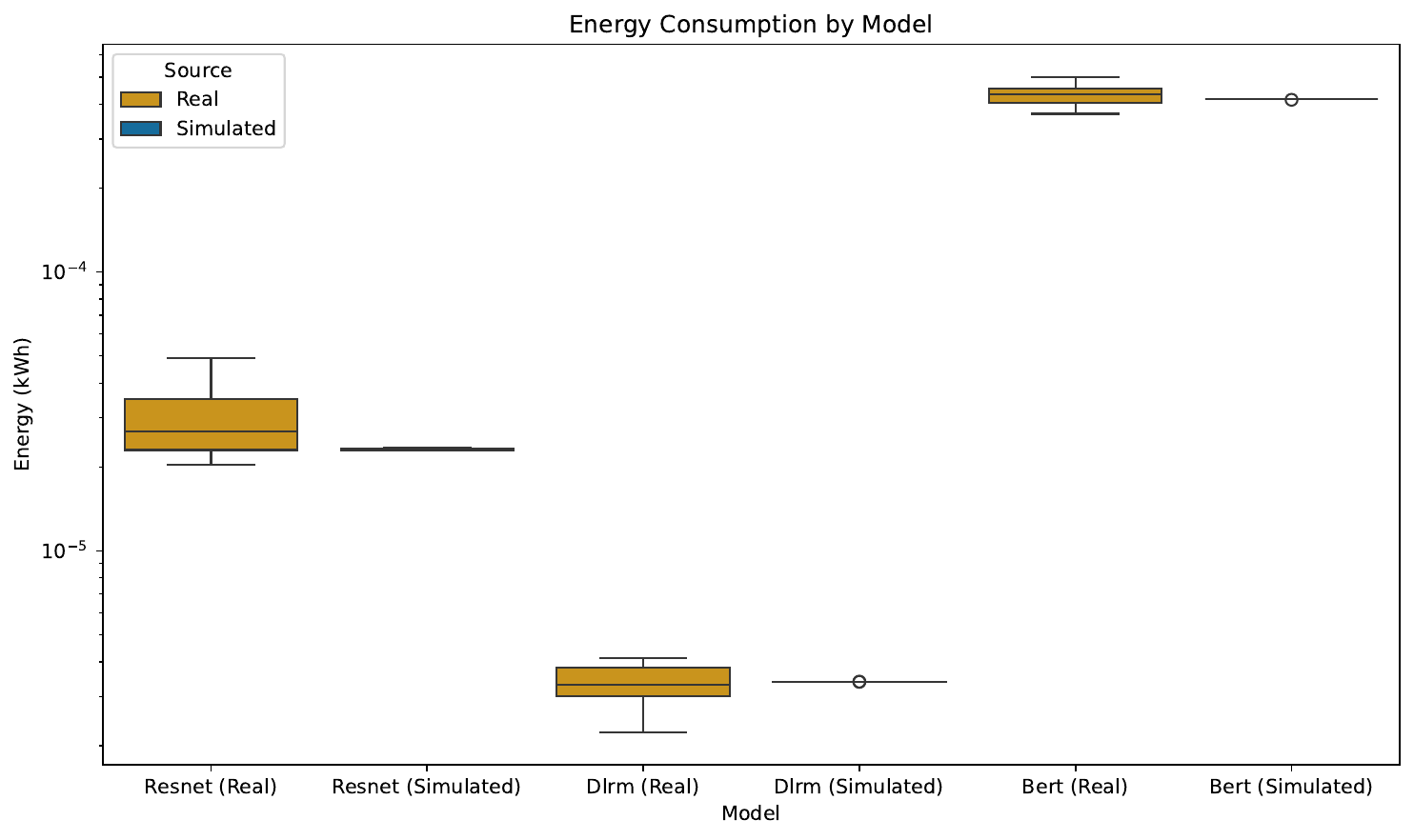}
    \caption{Energy consumption (\unit[mode=text]{\kWh}) by model, comparing real and simulated data. Log scale in the $y$ axis.}
    \label{fig:energyboxp}
\end{figure}

\begin{figure}[ht]
\centering    \includegraphics[width=.8\textwidth]{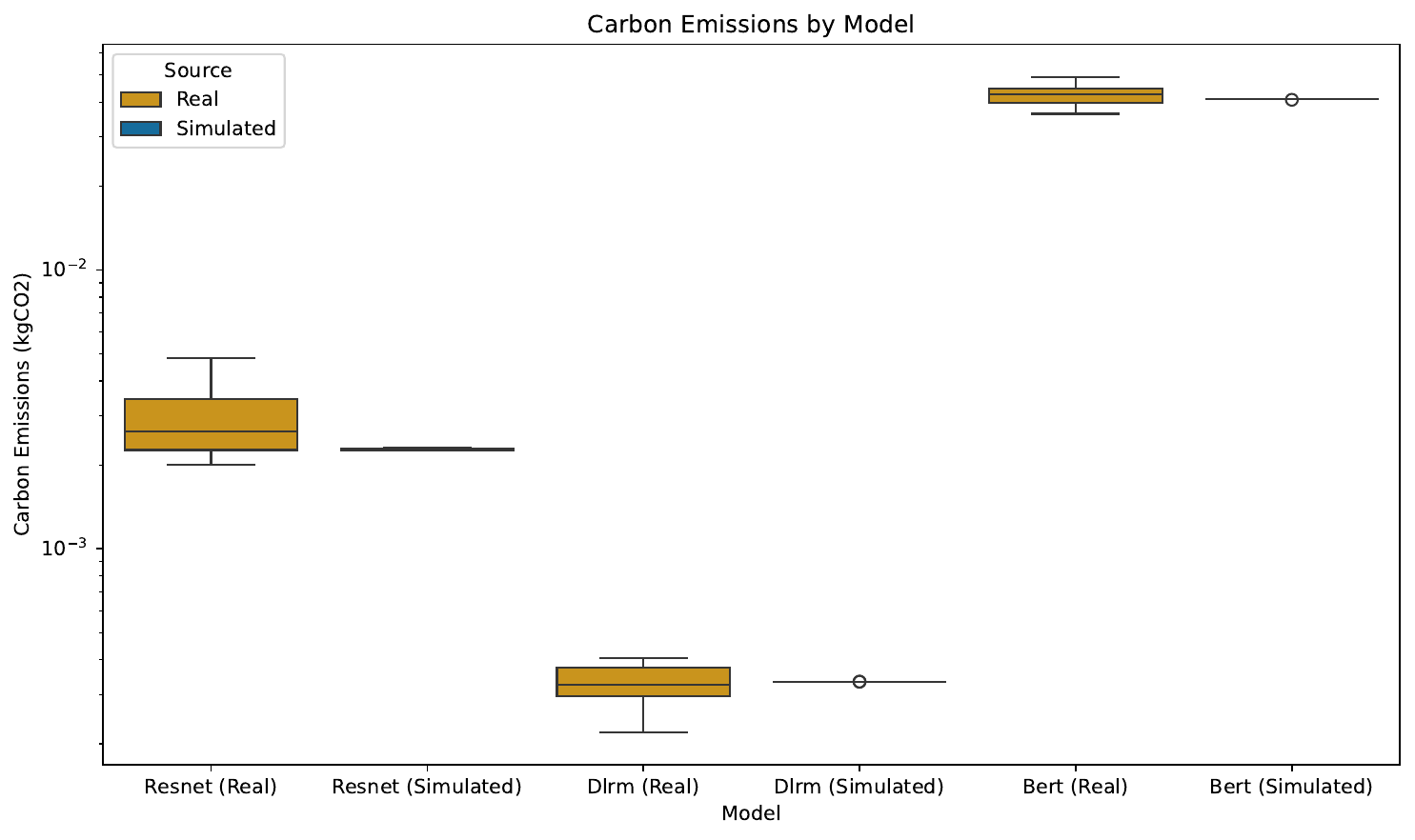}
    \caption{Carbon emissions (\unit[mode=text]{\kg} \ce{CO2}) by model, comparing real and simulated data. Log scale in the $y$ axis.}
    \label{fig:carbonboxp}
\end{figure}

The boxplots in Figure~\ref{fig:energyboxp} and Figure~\ref{fig:carbonboxp} visually illustrate the patterns observed in the quantitative metrics (MAPE and R\textsuperscript{2}). It is observed that, for all models, the simulated values show low variability, with boxes that are nearly collapsed and no presence of outliers. This reflects the deterministic nature of the Batsim simulation, which, by assuming constant consumption per power state, generates very stable results across executions.

In the case of ResNet18, there is a marked difference between the dispersion of real and simulated data, both in terms of energy consumption and carbon emissions. The real data present a wider interquartile range, indicating greater variation between executions, which aligns with its short runtime and susceptibility to system noise (such as frequency fluctuations and background processes). The median of the simulated values is also visibly lower than the real median, suggesting an underestimation of the energy load by the simulation model.

For DLRM, a behavior similar to that of ResNet18 is observed. Although overall consumption is even lower, the dispersion in real data remains evident, and the gap between the real and simulated medians is slightly smaller than in ResNet18, which is consistent with the intermediate mean absolute percentage error previously reported (MAPE $\approx 16\%$). The presence of outliers in the real data further reinforces the influence of external factors in short executions.

In the case of BERT Large, the boxplots reveal a closer match between real and simulated values. Both the medians and interquartile ranges are similar, especially in the energy consumption graph. Although there is still some variation in the real data, it is significantly smaller in relative terms, given the larger magnitude of the consumption. These visual results support the low MAPE (8.32\%) and reinforce that the simulation model is more reliable for heavy workloads and longer durations, in which the impact of system noise is mitigated.

The results indicate that the simulation-based approach is suitable for estimating average energy consumption and carbon emissions, especially for more intensive and longer-running workloads such as BERT Large. However, there are limitations for lightweight and fast models, since small variations in the real environment significantly affect measurements, while the simulation keeps values constant. More refined adjustments to the power states (\texttt{wattage\_per\_state}) or the introduction of dynamic load models could improve the accuracy of the simulation, bringing it even closer to real execution.

\section{Conclusion}

This work introduced a carbon footprint plugin developed in SimGrid and integrated into Batsim, enabling the simulation and monitoring of \ce{CO2} emissions based on the energy consumption of computing hosts. By allowing per-host configuration of carbon intensity and logging emissions during key scheduling events, the plugin extends Batsim's capabilities to support more environmentally aware simulations.

Through experiments comparing real and simulated executions of machine learning inference workloads, we demonstrated the plugin's ability to reproduce energy and carbon emission profiles with reasonable accuracy. These results indicate the plugin's potential as a valuable tool for researchers and practitioners who want to evaluate the environmental impact of task scheduling strategies and resource allocation policies in large-scale systems.

Future work includes improving simulation accuracy by refining energy models and validating the plugin across various scenarios. Furthermore, efforts will be directed toward making this plugin an official SimGrid extension, fostering long-term maintenance, and encouraging broader adoption by the research community.

\section{Acknowledgments}

This project was partially funded by FAPESP (procs. 23/00811-0, 24/09487-4, and 24/23163-7), by the Science for Development Center (CCD) `Carbon Neutral Cities' (FAPESP 24/01115-0), and by the CNRS International Research Center (IRC) `Transitions'. 

\bibliographystyle{sbc}
\bibliography{sbc-template}

\end{document}